\documentclass[dvips]{article}
\usepackage{astrosym}
\usepackage{icrctc07}

\title{Radio Detection of Ultra-High Energy Cosmic Rays$^1$}
\shorttitle{Radio Detection of UHECR}

\authors{Heino Falcke$^{1,2}$ for the LOPES collaboration}
\shortauthors{H. Falcke}
\afiliations{$^1$Department of
  Astrophysics, Institute for Mathematics, Astrophysics and Particle Physics, Radboud University,
  P.O. Box 9010, 6500 GL Nijmegen, The Netherlands\\ 
$^2$ASTRON, Oude
  Hoogeveensedijk 4, 7991 PD Dwingeloo, The Netherlands 
}
\email{h.falcke@astro.ru.nl}

\abstract {The radio technique for the detection of cosmic particles
has seen a major revival in recent years. New and planned experiments
in the lab and the field, such as GLUE, Anita, LUNASKA, Codalema,
LOPES as well as sophisticated Monte Carlo experiments have produced
a wealth of new information and I review here briefly some of the main
results with the main focus on air showers. Radio emission of
ultra-high energy cosmic particles offers a number of interesting
advantages. Since radio waves suffer no attenuation, radio
measurements allow the detection of very distant or highly inclined
showers, can be used day and night, and provide a bolometric measure
of the leptonic shower component. The LOPES experiment has detected
the radio emission from cosmic rays, confirmed the geosynchrotron
effect for extensive air showers, and provided a good calibration
fomula to convert the radio signal into primary particle
energy. Moreover, Monte Carlo simulations suggest that also the
shower maximum and the particle composition can be measured. Future
steps will be the installation of radio antennas at the Auger
experiment to measure the composition of ultra-high energy cosmic rays
and the usage of the LOFAR radio telescope (and later the SKA) as a
cosmic ray detector.  Here an intriguing additional application is the
search for low-frequency radio emission from neutrinos and cosmic rays
interacting with the lunar regolith. This promises the best detection
limits for particles above $10^{21}$ eV and allows one to go
significantly beyond current ground-based detectors. }

\begin{document}
\maketitle
\section{Introduction}
\footnotetext[1]{To appear in: 30th ICRC, Merida, Mexico 2007, Rapporteur Volume, ed.~J. F. Valdes- Galicia et al.}
Radio astronomy has always been closely connected to cosmic ray
physics. Already the very first cosmic radio emission detected by Carl
Jansky in 1932 originated from cosmic ray constituents in the Milky
Way. We now know that at low radio frequencies the diffuse Galactic
radio emission is mainly produced through synchrotron radiation of
relativistic electrons. They are propagating through the interstellar
medium and the Galactic magnetic field and were most likely
accelerated in supernova explosions.

Also, the brightest radio sources discovered thereafter, like quasars
and radio galaxies (active galactic nuclei) and supernova remnants,
are today the main suspects for the origin of cosmic rays. So, without
the advent of radio telescopes, we probably would not know much about
the non-thermal universe today.

It is therefore no surprise that radio antennas were early on also
considered for directly detecting cosmic ray air showers. In fact the
huge Lovell radio telescope in Jodrell Bank was initially built in
order to detect cosmic ray radar reflection
\cite{BlackettLovell1941,Gunn2005} --- it did not succeed but detected
the radar reflection of Sputnik instead and made history.

Radio detection of air showers has a number of advantages: the
detector material itself, a simple wire, is cheap, radio emission is
not absorbed in the atmosphere and can thus see the entire shower, and
interferometric techniques should allow relatively precise
localization. But does this work in practice and how does the radio
signal actually look like?

I will here mainly summarize some of the main recent results and not
recall the entire history of this field. Here one can point to the
well-known review by Allan from 1971 \cite{Allan1971} and a brief
summary of the early results given by Falcke \& Gorham
\cite{FalckeGorham2003}. Main points were the prediction of radio
Cherenkov radiation by Askaryan \cite{Askaryan1962a,Askaryan1965} and
the discovery of CR related radio pulses through Jelley et al. in 1965
\cite{JelleyFruinPorter1965} (a nice historical recount of the
discovery was given by Trevor Weekes \cite{Weekes2001}).

Despite many experimental problems at the time, quite a number of
basic properties of air shower were established within a decade
culminating in the empirical ``Allan formula'' \cite{Allan1971}.

A long hiatus of this field began in the 1970's, as witnessed by a
quote from Alan Watson in his 1975 ICRC rapporteur talk in Munich,
where he observed that ``Apart from work at 2 MHz which is planned for
Yakutsk, it is clear that experimental work on radio signals has been
terminated elsewhere.''

Occasional attempts with single or few radio antennas at EAS-TOP and
CASA/MIA \cite{GreenRosnerSuprun2003} did not lead to further radio
detections in the 1990's, making some colleagues (unjustifiably) even
doubt the reality of the earlier results in private conversations.

Only in recent years, the technique has seen an astounding revival. A
good overview is probably found in references
\cite{SaltzbergGorham2001,FalckeGorham2003,FalckeGorhamProtheroe2004,Nahnhauer2006}.

Scientifically, this started with attempts to detect radio emission
from neutrinos hitting the moon by Hankins, Ekers, \& O'Sullivan
\cite{HankinsEkersOSullivan1996} and Gorham et
al. \cite{GorhamHebertLiewer2004}.  For air showers, the realization
that the emission can be understood as geosynchrotron emission by
Falcke \& Gorham \cite{FalckeGorham2003} and Huege \& Falcke
\cite{HuegeFalcke2003} also inspired new efforts.

Technologically, the revival is certainly due to high-dynamic range
digital radio receivers and post-processing capabilities that are now
available. There is also a general revival in low-frequency radio
astronomy as seen in a number of projects such as
LOFAR\cite{FalckevanHaarlemdeBruyn2007},
MWA\cite{BowmanBarnesBriggs2007}, LWA\cite{KassimClarkeCohen2006} and
GMRT\cite{SwarupAnanthakrishnanKapahi1991}.

In the following we will give a summary of some of the results in this
field, with particular emphasis on radio emission from air showers and
results obtained with LOPES.

\section{Theory}
After the experimental realization in the late 1960's that the Earth
magnetic field is a factor in radio air shower emission, early
theoretical modeling by Kahn \& Lerche considered the Lorentz boosted
lateral current induced by the geomagnetic field
\cite{KahnLerche1966}, an approach that has been revisited very
recently by Werner \& Scholten
\cite{WernerScholten2007,ScholtenWernerRusydi2007}.

A different approach was presented by Falcke, Gorham, and Huege
\cite{FalckeGorham2003,HuegeFalcke2003} where the radio emission was
explained in terms of ``geosynchrotron emission''. This approach takes
an important extra factor into account, namely the curvature of the
trajectories of the individual electron/positron pairs in the
geomagnetic field. The fact that the emission region is smaller than a
wavelength (and optically thin) allows a relatively simple coherent
addition of the radio waves.  This ``single-particle approach'' makes
it straightforward to combine the radio emission with Monte Carlo
calculations \cite{HuegeFalcke2005a,HuegeUlrichEngel2007a}. 
The overall level of the geosynchrotron component seems to be
sufficient to explain the bulk of the observed radio emission
\cite{HuegeFalcke2005b}.  After all, geosynchrotron subsumes most of
the ``lateral current interpretation''. 

However, also in the geosynchrotron picture a couple of extra effects
still need to be taken into account, such as the current induced by
the change in charges through creation, annihilation and
recombination\cite{ScholtenWernerRusydi2007,Luo2006}. The
recombination of electrons may also lead to an additional
Bremsstrahlung component \cite{Luo2006}.  Also, the non-zero
refractive index of air and the original Askaryan effect through
Cherenkov emission from the charge excess
\cite{Engel2005,Meyer-VernetLecacheuxArdouin2007} will play a role at
some level. The static Coulomb contribution should also be looked at
\cite{Meyer-VernetLecacheuxArdouin2007} as well as optical depth
effects that could become relevant for air showers around $10^{20}$ eV
(proposed in the context of radio radar experiments \cite{Gorham2001}).

Hence, while major progress has been made, there is still room for
improvement. Nonetheless, the predictive power of current Monte Carlo
codes, if coupled with air shower simulations, is probably already
quite significant. This requires knowledge of the lepton evolution in
the showers and adequate shower libraries
\cite{LafebreHuegeFalcke2007}. 

\begin{figure}
\begin{center}
\noindent
\includegraphics[angle=270,width=0.475\textwidth]{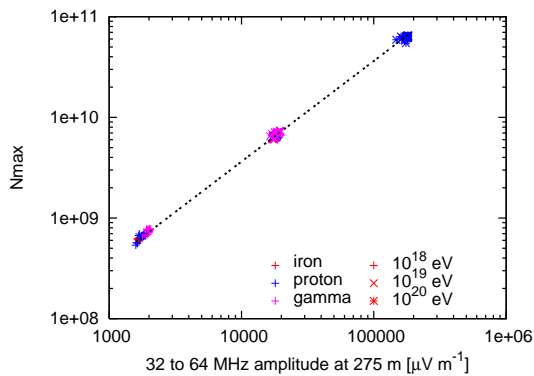}
\end{center}
\caption{Integrated 32-64 MHz radio flux measured at 275 m from REAS2 simulations as function of energy for different primary particles and showers with 60$^\circ$ zentith angle. Shower-to-shower fluctuations are only 5\%. Figure taken from reference \cite{HuegeUlrichEngel2007b}.}\label{fig01}
\end{figure}

\begin{figure}
\begin{center}
\noindent
\includegraphics[angle=270,width=0.475\textwidth]{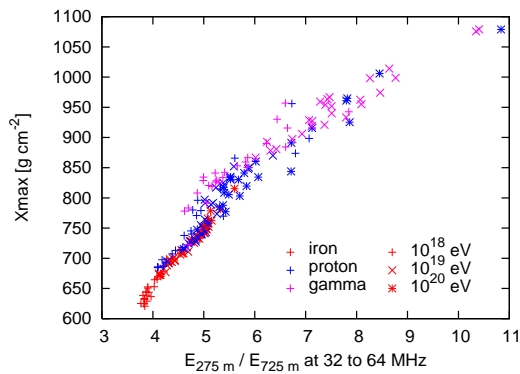}
\end{center}
\caption{Location of shower maximum, X$_{\rm max}$, vs. lateral slope
of radio emission from REAS2 simulations. The tight correlation
suggests that perhaps X$_{\rm max}$ should be measurable with radio
antennas and showers with 60$^\circ$ zentith angle. Figure taken from
reference \cite{HuegeUlrichEngel2007b}.}\label{fig02}
\end{figure}

Calculations with the REAS2 code and CORSIKA code by Huege et al.~have
recently shown some interesting results \cite{HuegeUlrichEngel2007b}:
If measured at a characteristic radius of $\sim300$ m from the shower
core the radio signal is tightly correlated with the primary particle
energy (Fig.~\ref{fig01}). Shower-to-shower fluctuations and different
elemental composition of the primaries induce just 5\% variations in
the radio flux. This is due to the fact that the radial radio
distribution on the ground pivots around a few hundred m for different
X$_{\rm max}$, depending a bit on shower geometry. This also means
that measuring the radial slope of the radio emission should give
clues for the location of the shower maximum (Fig.~\ref{fig02}) and
together with the absolute radio flux allow one to separate primaries
of different elemental composition. This tantalizing prediction
naturally requires experimental confirmation but already shows how
important the theoretical work is.


\section{LOPES} 
Quite a few experiments have been built or have been discussed in the
last couple of years that employ the radio detection method. We will
try to briefly discuss them here in turn.

A very productive experiment has been the LOFAR PrototypE Station
(LOPES), which made use of early prototype hardware developed for the
LOFAR radio telescope (see below) and helped to bring about the current
renaissance in radio detection techniques\cite{FalckeApelBadea2005}.

LOPES \cite{HornefferAntoniApel2004,FalckeApelBadea2005} was a
collaboration of radio astronomers involved in LOFAR and the groups
involved in the KASCADE\cite{KlagesApelBekk1997} and KASCADE Grande
array\cite{NavarraAntoniApel2004}. The idea was to put a significant
amount of radio antennas -- allowing for interferometric measurements
-- near a well-developed air shower array. This facilitates a cross
correlation between conventional air shower measurements and radio
observations.

\begin{figure*}
\noindent
\includegraphics[angle=0,width=0.39\textwidth]{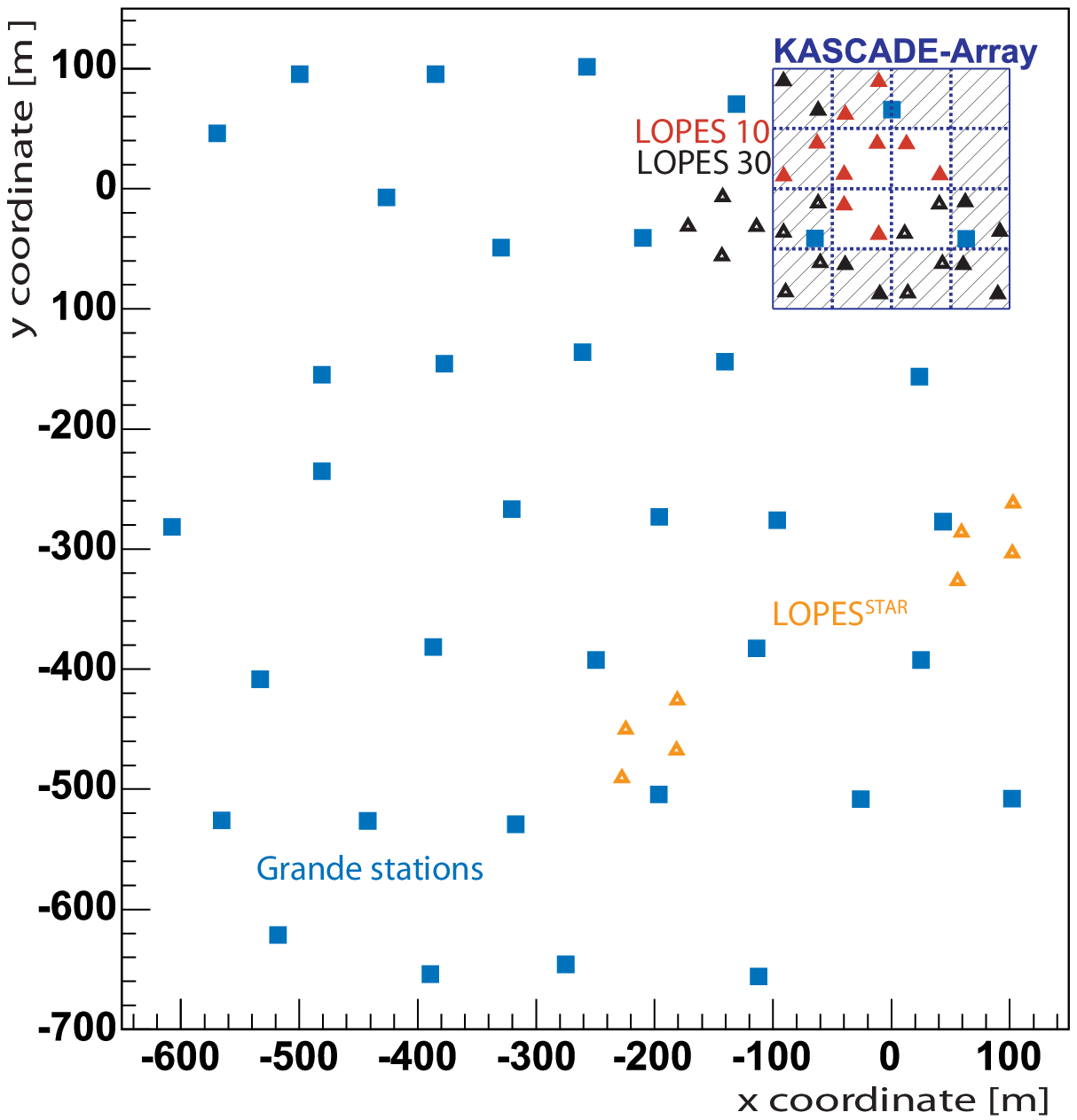}\vskip-5cm\hskip6cm\includegraphics[angle=0,width=0.6\textwidth]{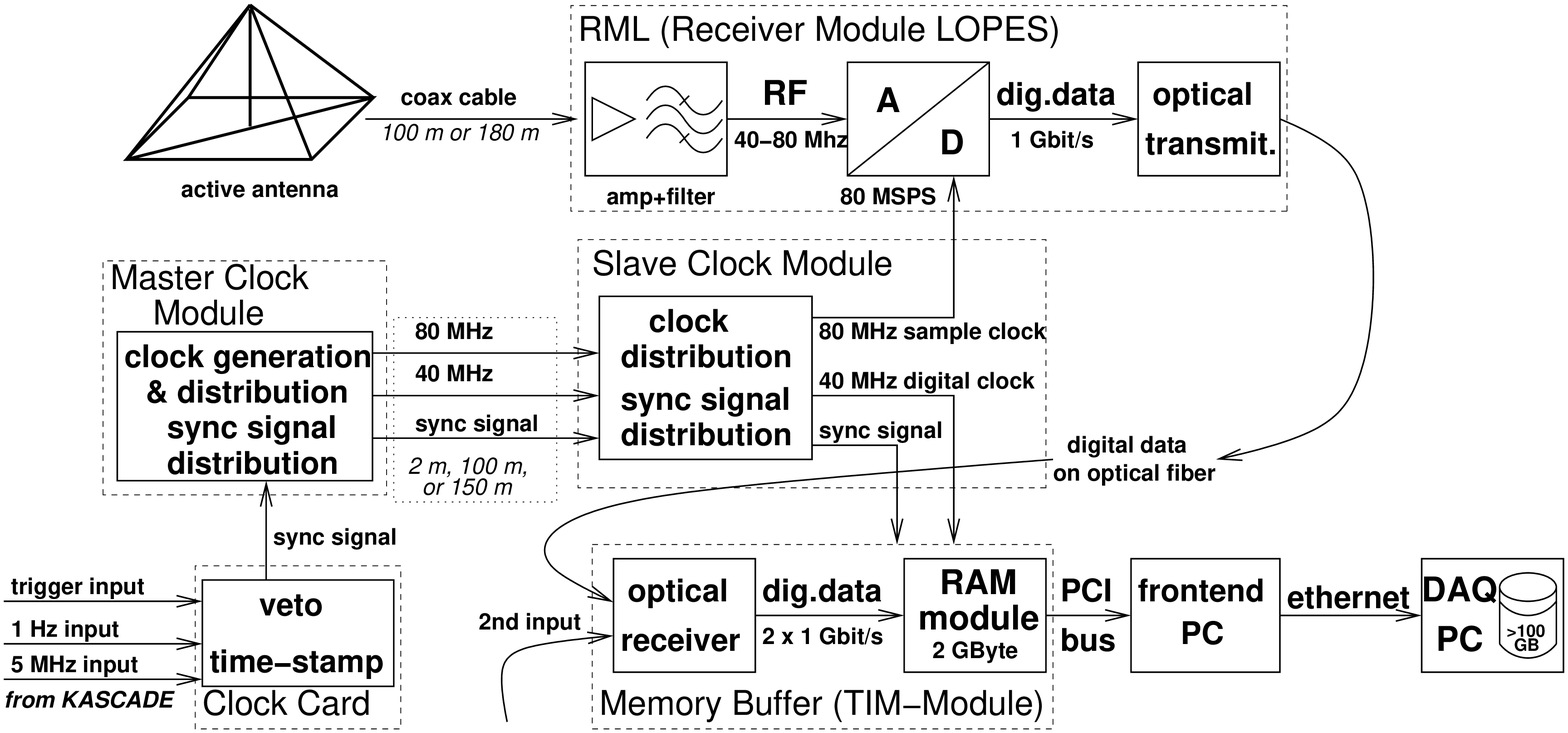}\vskip0.5cm
\caption{Left: LOPES and KASCADE Grande Layout. \label{fig03} Right: Schematic of LOPES Electronics\label{fig04}}
\end{figure*}

LOPES consisted initially of 10 single dipole antennas (LOPES10) that
were then expanded to 30 antennas (LOPES30). In the last phase LOPES
was again rearranged to have 20 antennas of which 10 are in a dual
polarization mode (Fig.~\ref{fig03}). Polarization investigations are
now underway (see Isar et al.\cite{IsarICRC2007}).

The LOPES antennas digitize the incoming radio waves with a 12 bit A/D
converter operating at 80 MHz. An analog filter restricts the
observable frequency range to 40-80 MHz, i.e.~the second Nyquist
zone. All antennas share a joint clock distribution and have
a 6.7 second ring-buffer which is triggered and read-out roughly twice
a minute by the KASCADE array (Fig.~\ref{fig04}). 

LOPES itself is restricted to the dimensions of KASCADE (200 m), but
with the help of KASCADE Grande events out to 500 m can be seen
\cite{ApelAschBadea2006}. 

The energies of cosmic rays seen in the radio at KASCADE is typically a
few times $10^{17}$ eV. Inclusion of KASCADE Grande provides
information up to $10^{18}$ eV. Below 10$^{17}$ eV the radio signal
vanishes in the noise.

Clearly, the Forschungszentrum Karlsruhe, where LOPES is located, is
not an ideal location of radio observations due to an enormous man-made
radio background noise (Radio Frequency Interference, RFI). This can
be overcome somewhat in the post-processing through digital filtering
methods.

Also, in addition to the LOPES antennas a few additional log-periodic
antennas (``Christmas trees'') with new electronics have been added in
order to develop a self-triggering algorithm \cite{AschICRC2007}. Here
the background noise is an even more severe problem. On the other
hand, studying self-triggering under these conditions, will allow one
to self-trigger almost everywhere else in the world as well.

\section{Codalema} 
Parallel to LOPES the CODALEMA experiment was
set-up\cite{ArdouinCharrier2006}, which initially had a complementary
approach: set up particle detectors near an existing radio astronomy
telescope at Nancay (France) and try self-triggering. This had the
advantage of a radio quiet site --- much better than LOPES --- and
well-calibrated antennas, but the disadvantage of having to calibrate
a new particle-detector array and to trigger on a yet not understood
radio signal. In the latest version the CODALEMA array is now
completely independent and employs a set of 16 wide-band active
dipoles aligned on two 600 meter long baselines in the North-South and
East-West directions. The radio array is triggered by a ground
detector array of 240 meters square containing 13 plastic scintillator
stations. The recording bandwidth is 1-200 MHz, but signal detection
is mainly done around 50 MHz, where also LOPES operates. Hence, the
two experiments are now quite compatible.

\section{LOFAR}

A next big step in radio detection of air showers will be the LOFAR
array which was planned as a large radio astronomy experiment
\cite{FalckevanHaarlemdeBruyn2007,RottgeringBraunBarthel2006}. In the
summer 2007 the project had to be downsized due to financial
shortfalls, however, it will still be a major step forward. According
to the current plans, LOFAR will consist of 36 antenna fields
(``stations'') in the Netherlands plus a number of stations across
Europe (E-LOFAR: Germany, UK, France, Sweden, Italy, Poland,
Ukraine). The first 20 stations are expected to operate early 2009.

\begin{figure}
\begin{center}
\noindent
\includegraphics[angle=0,width=0.475\textwidth]{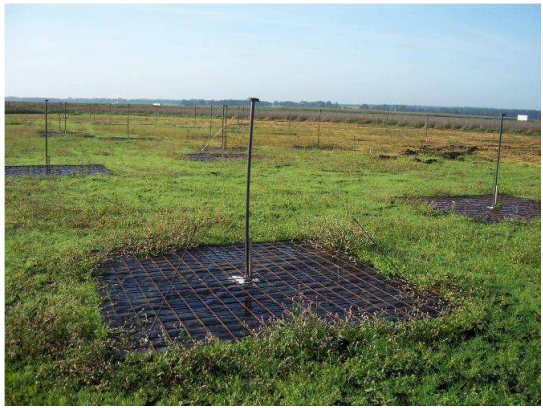}
\end{center}
\caption{LOFAR Low-Band Antennas}\label{figLBAs}
\end{figure}

\begin{figure}
\begin{center}
\noindent
\includegraphics[angle=0,width=0.475\textwidth]{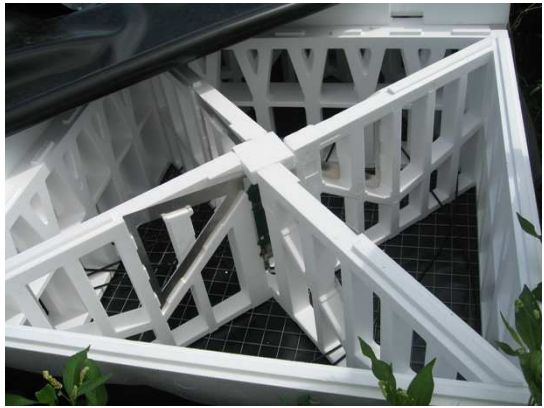}
\end{center}
\caption{Inside of a LOFAR High-Band Antenna (HBA) tile, showing one crossed bow-tie antenna.}\label{figHBA}
\end{figure}

Most of the antennas will be in a central concentration (``core'') of
2 km diameter, where 18 stations are foreseen. Each station has two
sets of receiver systems operating from 10-90 MHz (low-band antennas,
LBA, Fig.~\ref{figLBAs}) and 110-240 MHz (high-band antenna tiles,
HBA). The LBA field consist of 48 dual-polarization inverted-V
antennas, while the HBA fields consist of two sub-fields of 24
dual-polarization tiles. Each tile consists again of 16 bowtie-shaped
fat dipoles (Fig.~\ref{figHBA}).

This means that in the inner 2 km there will be more than 800
dual-polarization low-frequency antennas and about $\sim14,000$
high-frequency antennas ($>800$ tiles) that will be able to observe
bright radio events and deliver unprecedented detailed information
about radio shower properties.  LBAs and HBAs share one receiver, so
each LBA/HBA pair cannot observe at the same time, however, it is
well-possible to have one half of the receivers observe at the
low-frequencies, while the other half observes at higher frequencies.

For normal radio astronomical observations the radio data from one
station is combined into one data-stream (``digital beam-forming'') to
look in a predetermined direction. However, every antenna is also connected
to a one second ring-buffer with FPGA-based processing and triggering
capability. This allows one to trigger on the raw radio data stream in
an intelligent way. Since 8 antennas share one memory board (with 4
FPGAs), there is the possibility to trade the number of antennas for
triggering power or buffer length.

\section{SKA} 
As a next step in radio astronomy at the low frequencies the Square
Kilometre Array (SKA) project is planned for $>2015$
(www.skatelescope.org), which will provide even more opportunities for
radio detection of cosmic rays and neutrinos
\cite{FalckeGorhamProtheroe2004}.  The SKA will employ a mix of
receptor technologies (dipoles, tiles, small dishes) depending on the
frequency range (70 MHz- 10 GHz) and have a phased roll-out. The first
phases will concentrate below 1 GHz and will be of high interest for
radio particle detection as discussed here.

\section{Auger Radio}
In the spirit of the LOPES experiment, putting radio antennas next to
existing cosmic ray experiments, it makes sense to also place radio
antennas at today's largest cosmic ray array, the Pierre Auger
observatory \cite{Watson2008,AugerCorrelation2007}. This has been
attempted recently with a few prototype radio antennas (van den Berg
et al.~\cite{BergAugerICRC2007}), including some LOPES antennas. The
medium-term goal is to cover a 20 km$^2$ region with self-triggering
radio antennas. In the same region an infill array and an upgrade to a
fluorescence telescope will lower the energy threshold of Auger. This
will nicely connect to the LOPES and CODALEMA measurements in energy
and allow for the first time triple coincidences between radio,
particle detectors, and fluorescence and hopefully further
dramatically increase the quality of the data.

\section{Radio in Ice and from the Moon}
Apart from the radio air shower experiments a high level of attention
has also been attracted by the possibility to detect showers generated
in solid media, such as ice, salt or the lunar regolith. This goes
back to the original suggestion by Askaryan
\cite{Askaryan1962a,Askaryan1965}. Further theoretical progress in the
1990's by Alvarez-Muniz and Zas et
al.~\cite{AlvarezMunizVazquezZas2000,ZasHalzenStanev1992,Beresnyak2003} and
successful accelerator experiments, validating the theory, breathed
new life into this field
\cite{SaltzbergGorhamWalz2001,GorhamBarwickBeatty2007}.

One idea is to use the huge detector volume of the moon and observe it
with sensitive ground-based radio telescopes in search for nanosecond
pulses which are dispersed by the Earth ionosphere.  First such
experiments were made with the Parkes radio telescope
\cite{HankinsEkersOSullivan1996} and the 64 m Kalyazin radio telescope
\cite{BeresnyakDagkesamanskiiZheleznykh2005}. An experiment (GLUE)
using the NASA deep space network antenna at Goldstone received wide
attention and produced interesting limits on the ultra-high-energy
neutrinos flux \cite{GorhamHebertLiewer2004}. Currently extensive
experiments, LUNASKA and NuMoon, are progressing at the Australia
Telescope Compact Array (ATCA, James et
al. \cite{JamesCrockerEkers2007}) and the Westerbork Synthesis Radio
Telescope (WSRT, Scholten et al. \cite{ScholtenICRC2007}). Later LOFAR
and the SKA could be used to further improve the current upper limits
to interesting levels or even detections
\cite{ScholtenBacelarBraun2006,FalckeGorhamProtheroe2004}.  Even a
detection of neutrinos using radio {\it on} the moon has been
considered\cite{JesterFalcke2008}.

Instead of looking up, one can also look down on the Earth and use for
example the large ice sheet of Antarctica or salt domes as detector
targets. While salt domes \cite{MilincicGorhamMiocinovic2007} are
currently not pushed very strongly due to high drilling costs,
experiments involving Antarctica are flourishing.

For quite some time already the RICE experiment
\cite{KravchenkoCooleyHussain2006} has radio antennas in the ice near
the IceCube array \cite{KarleAhrensBahcall2003} and a major extension
of IceCube with radio antennas is actively discussed
\cite{KarleAhrensBahcall2003,KarleICRC2007}. Alternatively radio
antennas have been tested at the Ross ice shelve in Antarctica
(ARIANNA) which is logistically more conveniently located
\cite{Barwick2006,BarwickAriannaICRC2007}.

An alternative approach to embedding a large number of radio antennas
in the detector volume is to just fly over it and use the long range
capability of radio detection. Gorham et
al.\cite{LehtinenGorhamJacobson2004} have used a military satellite
(FORTE) to search for neutrino induced radio pulses from the
ice. Recently the dedicated balloon experiment ANITA
\cite{BarwickICRC2007} was launched for the first time to circle
Antarctica and to detect there distant radio pulses from up-going
neutrinos. Unfortunately the flight was cut short by unfortunate wind
conditions, but nonetheless the data analysis is proceeding.

This brief summary already shows that the number and breadth of radio
experiments for cosmic ray and neutrino detection is rather large
already. 

\section{Experimental Results \& Calibration}
In the following we will summarize some of the important experimental
conclusions concerning the air shower radio properties that have been
found recently, here mainly focused on air showers and based on the
LOPES results.

\begin{figure}
\begin{center}
\noindent
\includegraphics[angle=0,width=0.475\textwidth]{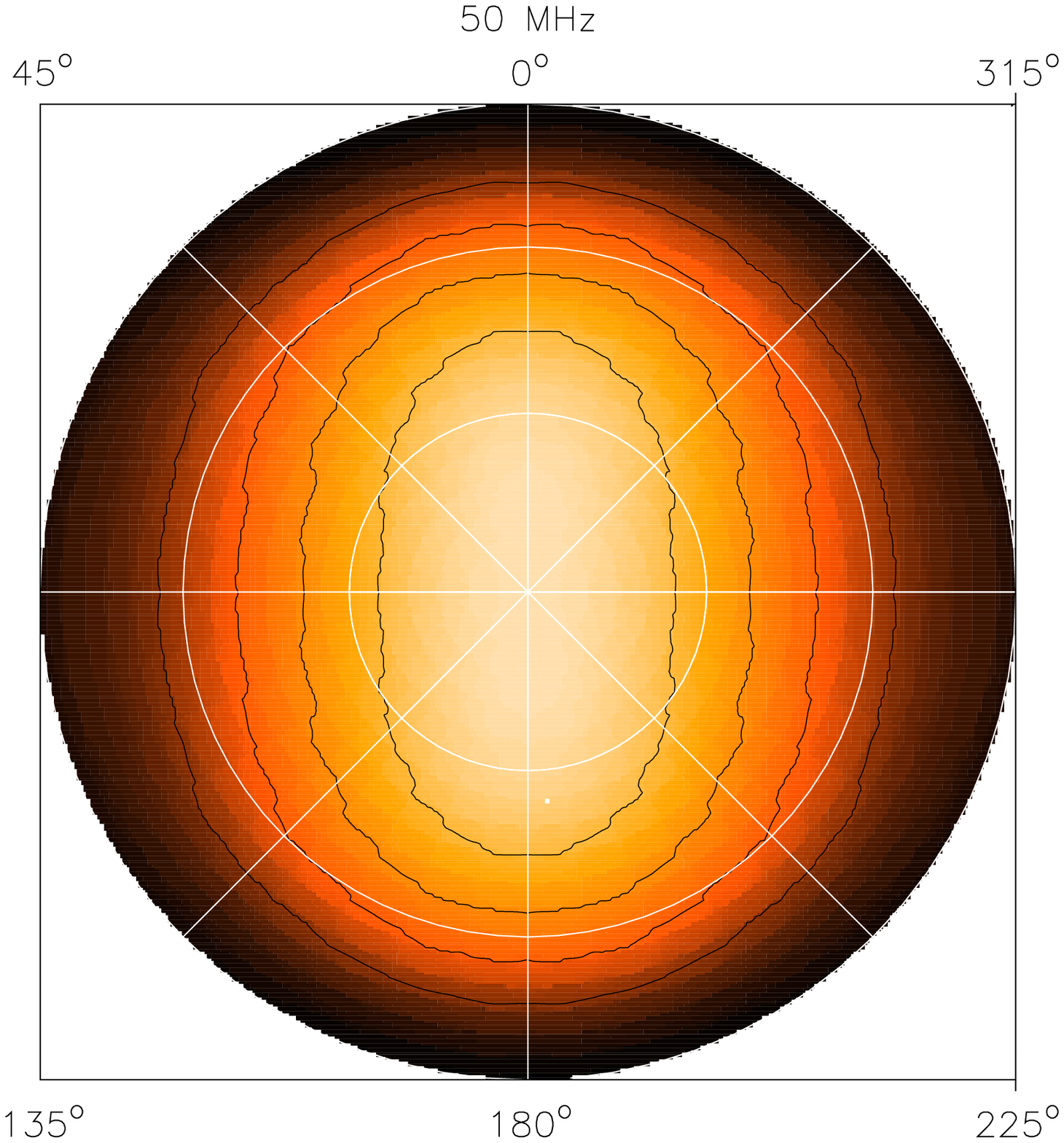}
\includegraphics[angle=0,width=0.475\textwidth]{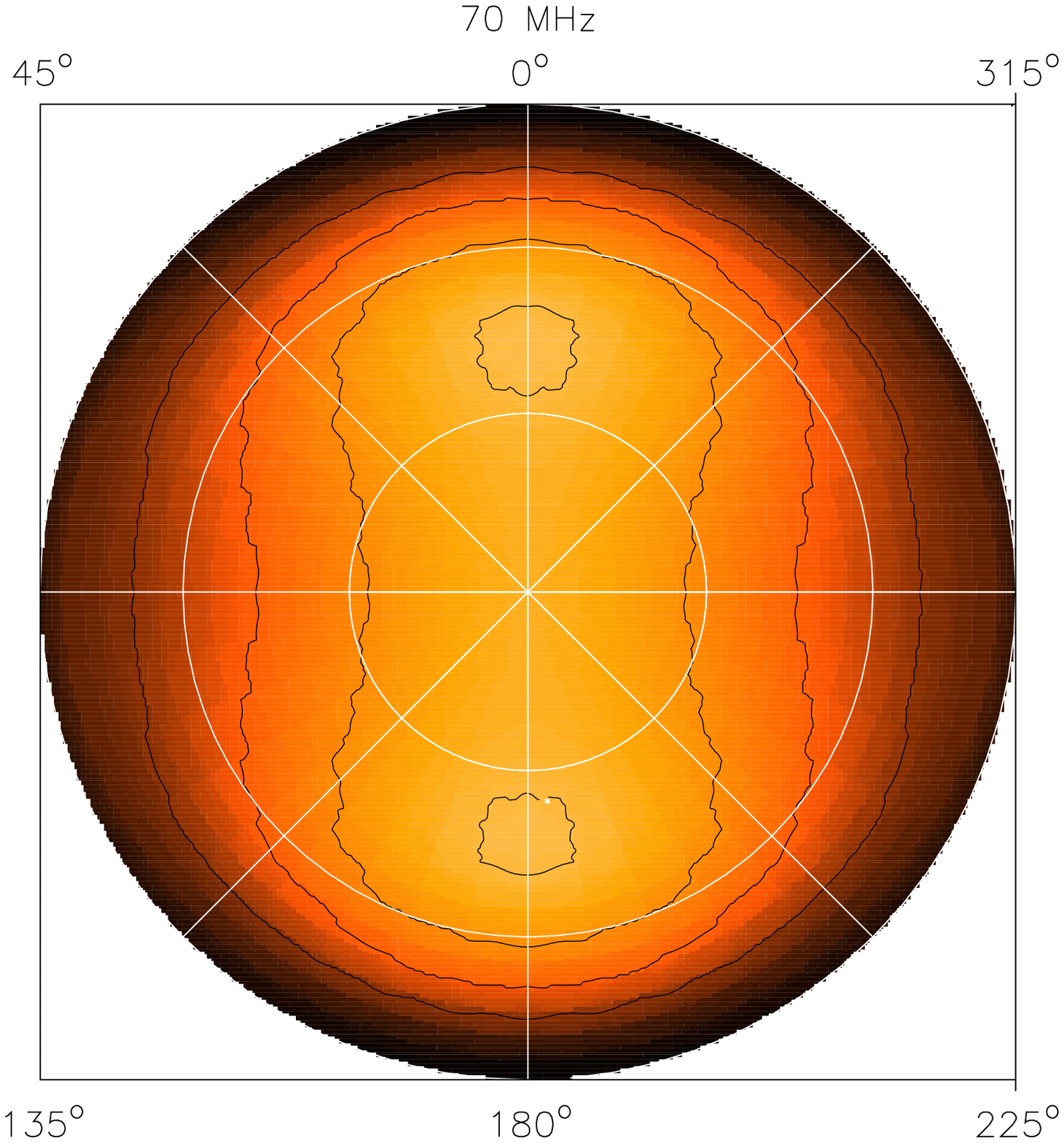}
\end{center}
\caption{Gain pattern of a LOPES antenna at 50 MHz (top) and 70 MHz (bottom) showing the entire sky. The color scale goes from a gain of 0 (black)to 5 (white).
\label{figgpat}}
\end{figure}

First of all one has to realize that the simplicity of the antenna
comes at the cost of more complicated calibration. The sensitivity of
a dipole depends on frequency, direction, and polarization. For LOPES
the absolute calibration has been performed as part of the PhD thesis
of S. Nehls \cite{NehlsHakenjosArts2008} using an elevated calibrated
reference antenna. On the other hand, given the simple structure of
the antenna, it is also possible to calculate the expected beam
pattern on the sky using standard antenna simulation packages.  

An example is shown on Fig.~\ref{figgpat}, which shows that the beam
shape is elongated and even not peaking towards the zenith for
frequencies above the resonance frequency of the LOPES dipoles ($\sim$
60 MHz). The elongated structure is related to the orientation of the
dipoles (EW) and would be rotated by 90$^\circ$ for the other (NS)
polarization. In turn this also means that the crossed-dipole, if
uncalibrated, will always produce highly polarized signals.

\section{Energy Calibration}
\begin{figure}
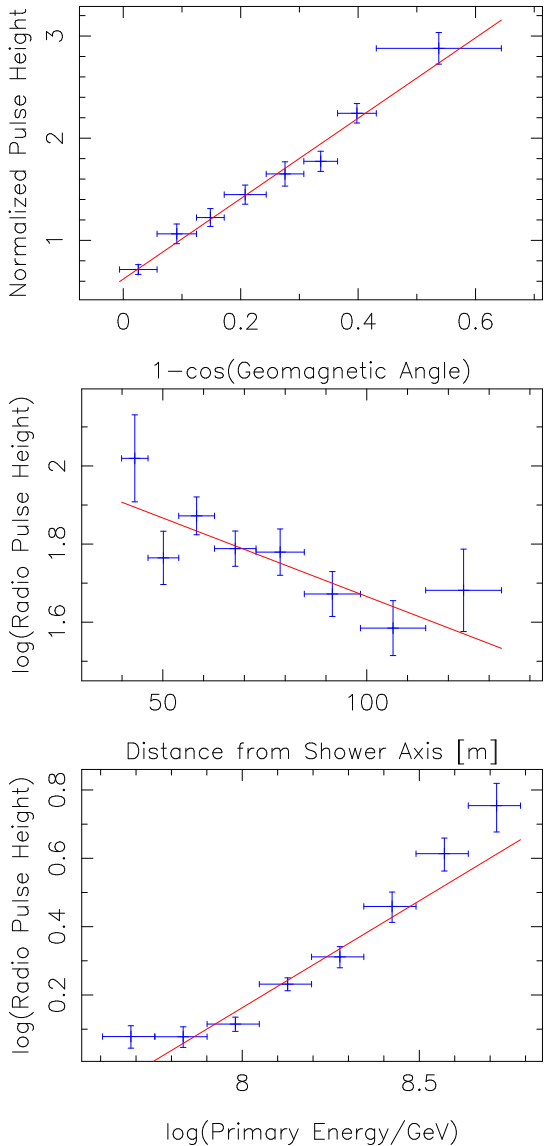

\begin{center}
\noindent
\includegraphics[angle=270,width=0.475\textwidth]{icrc1317_fig-geomag.ps}
\includegraphics[angle=270,width=0.475\textwidth]{icrc1317_fig-distance.ps}
\includegraphics[angle=270,width=0.475\textwidth]{icrc1317_fig-energy.ps}
\end{center}
\caption{Calibration results of LOPES showing the normalized radio
voltage vs. air shower parameters where the other parameters have been
divided out. Each point represents the average and spread of all
events in that bin. Panels a-c are from top to bottom.
\label{figlopcal}}
\end{figure}

In addition to the antenna dependencies, the radio emission will also
depend on the shower geometry. The main factors that have been
identified are the particle energy, $E_{\rm p}$, the angle between
shower axis and geomagnetic field (``geomagnetic angle''), $\alpha$,
the zenith angle, $\theta'$, and the distance from the shower core,
$r$. For an east-west polarized antenna one finds that the radio
emission for showers at the same energy and distance is proportional
to $1-\cos\alpha$ (Fig.~\ref{figlopcal}a), the signal drops
exponentially with radius (Fig.~\ref{figlopcal}b), and increases
linearly with primary particle energy (Fig.~\ref{figlopcal}c).

Altogether this has been nicely parametrized in Horneffer's formula
\cite{HornefferICRC2007} (at the moment valid only for the EW polarization):
\begin{eqnarray}
&\epsilon_{\rm est}  =
(11\pm1.)
\left((1.16\pm0.025)-\cos\alpha\right) \cos\theta & \nonumber \\
& \exp\left(-\frac{r}{\rm (236\pm81)\,m}\right)
\left(\frac{\rm E_{p}}{\rm 10^{17}eV}\right)^{(0.95\pm0.04)} 
\left[\frac{\rm \mu V}{\rm m\,MHz}\right]&
\label{eq:horneffer-energy}
\end{eqnarray}

We note that the exponential decay of the radio signal is also seen by
the CODALEMA experiment \cite{ArdouinCharrier2006}. 

This prescription can now be inverted to predict the energy of the
incoming particle. Comparison between the energy predicted from radio
with the energy estimated from KASCADE Grande, shows a scatter of
27\% between the two methods for $E_{\rm p}>10^{17}$ eV. This is very
encouraging, given that shower-to-shower fluctuations in the KASCADE
Grande estimate alone should produce a 25\% scatter. Hence, the
scatter in the radio measurements should be much less.

This would support the claims from Monte Carlo simulations
\cite{HuegeUlrichEngel2007b} that radio is a good tracer of the
energy. The main reason why one suspects lower scatter in the radio
measurements with respect to particle detection on the ground is the
fact that radio emission in not absorbed in the atmosphere. Hence,
radiation from every particle is visible on the ground --- it is in
that sense a bolometric measurement. Variations in the location of the shower
maximum will be less dramatic compared to measurements of particles on
the ground which are just a fractional tail of a quickly declining
function, whose values are quite sensitive to X$_{\rm max}$.

\section{Spectrum}
One topic that had been difficult to tackle in the past has been the
spectral shape of the radio signal, i.e.~how much power is emitted at
which frequency? Historic experiments were relatively narrow band and
non-simultaneous data had to be combined. Modern broad-band receivers
allow one to study the instantaneous spectral index, but require
careful bandpass calibration. 

\begin{figure}
\begin{center}
\noindent
\includegraphics[angle=0,width=0.475\textwidth]{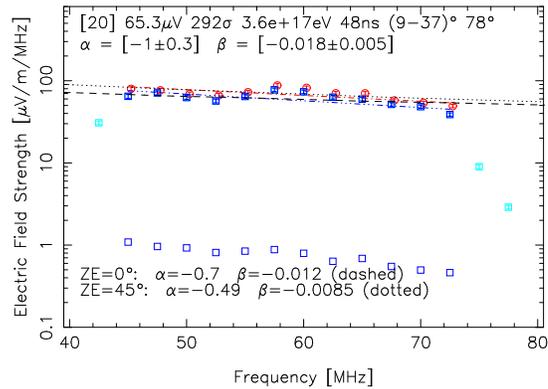}
\end{center}
\caption{Example spectrum of a LOPES event for two different methods
of beam-forming. The measured spectral slope is
$\epsilon\propto\nu^{-1}$.  The lines represent two spectra from Monte
Carlo simulations. Also shown is the noise spectrum that has been
corrected. \label{figspectrum}}
\end{figure}

Nigl et al.~\cite{NiglSpectrum2008} employed two methods to get the
spectral shape (Fig.~\ref{figspectrum}): Fourier transform of the (not
squared) electric field around the pulse position and measurements of
the pulse heights after applying narrow-band digital filters to the
signal. Both methods give consistent results.

The spectra can be represented by a power-law function or an
exponential decay. For the narrow frequency range of LOPES, extending
only a factor of two, we cannot distinguish between the two
prescriptions. For a power-law function the average spectral index is
$\nu^{-1\pm0.3}$. This would mean that the power of the signal falls
of with $\nu^{-2}$. This is consistent with the simple expectations of
coherent geosynchrotron \cite{FalckeGorham2003} and only slightly
steeper than the Monte Carlo simulations suggest
\cite{HuegeFalcke2005b}.  Also, the Codalema experiment finds powerlaw
spectra with spectral indices in the range -1.5 to 0. LOPES sees
spectral indices in the range -1.5 to -0.4, so there is some
agreement, but more detailed investigations have to be performed in
the future.

\section{Direction \& Imaging} 
The next question then is, how well can we localize the radio
emission? This has become of particular importance given the finding
of anisotropies and correlations between cosmic ray arrival directions
and nearby extragalactic objects \cite{AugerCorrelation2007}.

Using radio astronomical imaging techniques, we can actually image the
radio flash from the air shower. For LOPES L. B\"ahren has developed a
special tool (``skymapper'') which can actually do this on the tens
of nanosecond (i.e., the sampling rate) level (see
Fig.~{\ref{figskymap}) and in three dimensions (Fig.~\ref{fig3D}).

\begin{figure}
\begin{center}
\noindent
\includegraphics[angle=0,width=0.475\textwidth]{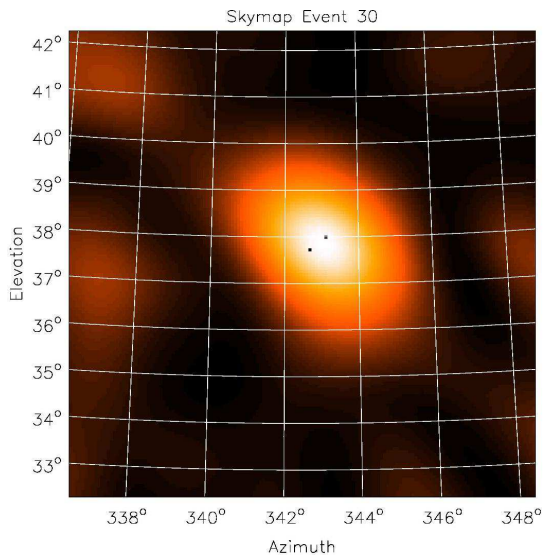}
\end{center}
\caption{Image of a bright radio flash at the time of maximum. The
dark dots mark the fitted LOPES position and the direction found by
KASCADE. The size of the ellipse is determined by the expected image resolution.
\label{figskymap}}
\end{figure}

\begin{figure}
\begin{center}
\noindent
\includegraphics[angle=0,width=0.475\textwidth]{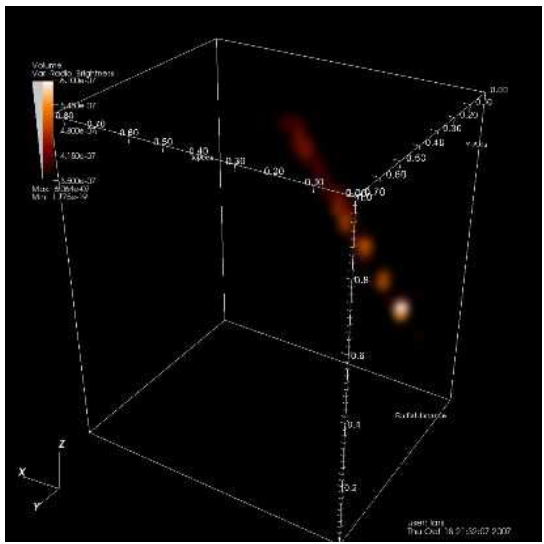}
\end{center}
\caption{ Three dimensional view of the radio emission from a cosmic
ray air shower detected with LOPES. 3D sidelobes have been ``cleaned''
by only displaying high-brightness regions. The peak at 3 km height
appears to be real, but the extended structure along the axis is
strongly affected by remaining sidelobes.
\label{fig3D}}
\end{figure}

This issue of the positional accuracy has then been further
investigated by Nigl et al. \cite{NiglDirections2008} using LOPES
data. Conventional interferometry is very sensitive to positional
changes. For point sources one expects an angular error
$\Delta\alpha_{min} = \pm\frac{1}{2}\frac{1}{\rm
SNR}\frac{\lambda}{D}$ in the azimuthal direction, where SNR is the
signal-to-noise ratio, $D$ the separation of the antennas, and
$\lambda$ the observing wavelength. For high SNR images the positional
accuracy for point sources is always better than the image
resolution. Hence, for an SNR of 10 for an antenna separation of 100 m
and observing frequencies around 60 MHz, as in LOPES, one expects an
error of only 0.15$^\circ$, while the point-spread function of the
interferometer has a much larger width of about 2-3$^\circ$ in
azimuth.

Comparisons of the shower direction (assuming a fixed shower core)
between KASCADE and LOPES, actually shows an average offset of
$1.3^\circ$ (Fig.~\ref{figoffsets}), which does decrease with
increasing signal level (and increasing SNR). This is not bad compared
to the imaging resolution of LOPES, but we would have expected better
results still. So, what is the dominating source of error?

\begin{figure}
\begin{center}
\noindent
\includegraphics[angle=0,width=0.475\textwidth]{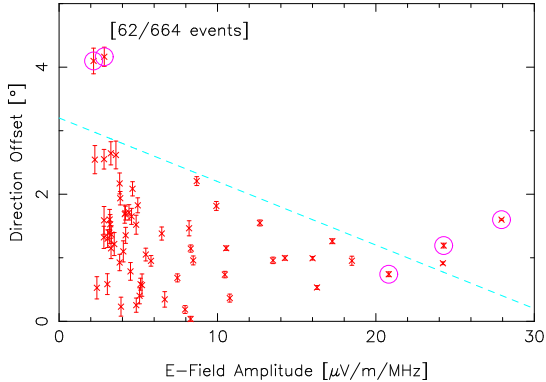}
\end{center}
\caption{Offsets on the sky between the direction of the air shower axis
as measured with KASCADE and LOPES vs the radio pulse amplitude. Events with circles were taken during thunderstorm conditions.
\label{figoffsets}}
\end{figure}

Some insight can be obtained from Fig.~\ref{figelv} which shows that
the location of the radio centroid on the sky is also
a function of the distance or radius of curvature.  One has to
remember that the shower maximum of the showers that LOPES sees is
just a few km high. This is still in the near field of the
interferometer. This means that radio waves emitted in the shower
maximum will not appear as a plane wave, but will show a curvature
with a radius corresponding to the distance from the observer. In the
real world, the wave front will be even more complicated, since the
emission is not constrained to a small region but extends along the
shower axis. In principle one has then a wavefront that is the
superposition of many spherical waves emitted at different positions.

In a proper 3D imaging process one would try to deconvolve the data
and put together a 3D image cube. We are not yet able to do this. So,
all we can do at present is to try different radii of curvature and
search for the maximum in the emission. This is what Fig.~\ref{figelv}
shows: different cross sections of a radio image focused at different
distances in steps of 250m. The maximum is found at a distance of
about 3 km. This is 30 times farther than the typical baselines on the
ground and only possible due to the good SNR. What is clear from the
figure is that not only the emission level changes but also the
position of the maximum. The problem is that for small radii of
curvature a small change in radius is similar to an inclination of a
plane wave. Inclining the (virtual) receptor plane implies a
positional shift on the sky. As seen in this example, the shift can be
up to 3$^\circ$. Hence, an error in the radius of curvature
determination or -- perhaps more important -- a non-spherical
wavefront will propagate into an positional error!

This requires that radio shower parameters are really derived from a 4
(or 6) dimensional data cube, consisting of time, 3 spatial
coordinates, and potentially 2 shower core location parameters
\cite{ApelAschBadea2006}.

So, any experiment will improve its spatial accuracy not only with
greater baselines but also with increasing the number of antennas. In
addition we need to understand the exact geometry of the radio shower
front. Here, further simulations and the detailed observations with
the many antennas of LOFAR should clarify that issue. This may make
further dramatic improvements in the astrometry of radio air showers
possible.

\begin{figure}
\begin{center}
\noindent
\includegraphics[angle=270,width=0.475\textwidth]{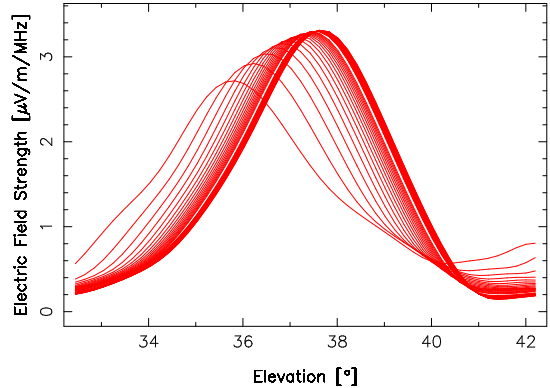}
\end{center}
\caption{Cross section along the elevation axis in an image of a cosmic
ray radio flash for different radii of curvature. The maximum field
strength is found around a radius of 3km.\label{figelv}}
\end{figure}

\section{Electric Fields and Lightning} 
One other important factor that has been looked at with some worries, is
the influence of the atmospheric electric field on the radio
emission. While the Earth magnetic field is very stable, the electric
field can change significantly from 1-10 V/cm during fair weather,
to 100 V/cm in heavy rain clouds (Nimbostratus), and up to 1000 V/cm in
severe thunderstorms. If the radio emission is affected by the
electric field one would not be able to interpret the radio signal
quantitatively, since measuring precisely the instantaneous electric
field structure is almost impossible.

To get a first idea of the importance of the electric field, we could
simply look at the Lorentz force, $F=e(\vec E+\vec v/c\times \vec B)$
in {\it cgs}, which is driving the geosynchrotron emission. For a
relativistic particle ($v/c\simeq1$) the electric field will then
dominate if $E>B\simeq 150 {\rm V\,cm^{-1}} (B/0.5\,{\rm G})$. Hence,
from this simple approximation one would expect only for severe
weather a modification of the radio signal.

Buitink et al.~\cite{BuitinkApelAsch2007} investigated radio pulse
heights with LOPES under different weather conditions. They selected
time slots where the local weather station recorded clear weather,
heavy rain, or thunderstorms and compared the three
(Fig.~\ref{figthunder}). Within the errors, the radio pulses followed
the previously found correlation with the geomagnetic angle (if
normalized to the same energy and radius). However, significant
outliers are found in the thunderstorm data set -- and only there. The
amplification of the radio signal can be a factor ten.

\begin{figure}
\begin{center}
\noindent
\includegraphics[angle=270,width=0.475\textwidth]{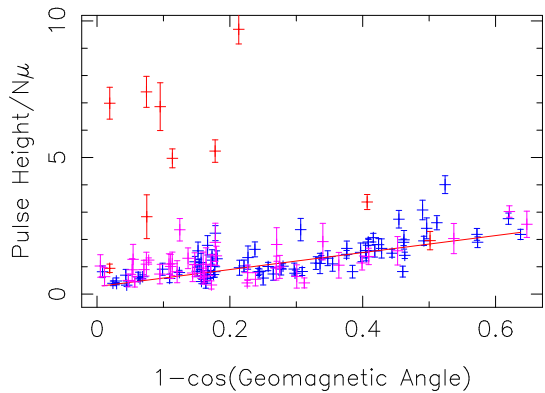}
\end{center}
\caption{Radio pulse height (normalized by energy) versus the geomagnetic angle. Blue dots are taken during clear weather, purple during nimbostratus, and red points during thunderstorm  conditions.\label{figthunder}}
\end{figure}

This seems to confirm the simple estimate we made above and the fact
that the Lorentz force is dominating. Further verification comes from
Monte Carlo simulations. Buitink et al. \cite{BuitinkICRC2007} have
now included electric fields in the CORSICA and REAS2 codes, allowing
one to model the E-field influence in detail. The simulations have
calculated radio pulses for different values of the electric
field. Again a significant amplification is seen as soon as the
E-field reaches 1000 V/cm, while at 100 V/cm the radio emission shows
only little differences.


The CORSIKA simulations also show a modification of the
electron/positron energy distribution and the shower structure. Many
pairs are deflected significantly from the shower axis, which might be
detectable with particle detectors.

Finally, we note that besides the radio pulse height also other
parameters are impacted by thunderstorm electric fields. For example,
in the investigation of the emitted radio spectrum one bright radio
event stood out with a much steeper radio spectral index. It was found
to be a thunderstorm event.

Moreover, also in the positional offsets between LOPES and KASCADE
thunderstorm events stand out. In Fig.~\ref{figoffsets} all outliers
are events measured during thunderstorms. Whether this is due to an
actual deflection or an asymmetry in the radio emission is not yet
clear.

In summary, we can state that air showers passing through the strong
fields of thunderstorm clouds are brighter, further offset from the
shower core measured on the ground, and have a steeper radio spectrum.
For the measurement of cosmic rays this means that radio --- at least
at frequencies above 40 MHz --- remains a reliable technique as along
as data taken during thunderstorm is discarded.

On the other hand, the current results strongly suggest that in high
E-fields not just the radio emission is altered, but the entire shower
(at least the electronic part). This point may warrant further
investigation for its own sake. Moreover, it has been speculated that
cosmic ray air showers could play a role in initiating lightning
through a runaway breakdown effect \cite{GurevichZybin2004}. 

Radio methods could help to investigate this connection
experimentally, since both -- the lightning strike and the air shower
-- would be detectable by the same instrument. Also, further Monte
Carlo simulations will investigate whether there is enough energy gain
through the electric field or ionization through the air shower to
actually start the runaway breakdown process. The LOFAR project will
try to address some of these issues.

\section{Self-triggering}
Overall, the current results have provided a very comprehensive picture
of the radio properties. However, whether the radio detection will
mature into a standard technique depends on whether it is possible to
actually trigger on the radio signal. This has been attempted but the
final breakthrough still stands out. Recent attempts were made for
example by the CODALEMA experiment
\cite{RavelDallierDenis2004}. Within LOPES, LOPES$^{\rm STAR}$
\cite{AschICRC2007} has been designed specifically to investigate this
issue in more detail.

We do now understand where the challenges lie. In many cases,
disturbing radio interference has actually been generated by the
devices and electronics of the cosmic ray experiments themselves. So,
designing radio quiet electronics, power supplies and data
communication is an important first step. Also, radio contains an
enormous amount of information (as witnessed by every FM receiver) and
many processes of the modern world produce wanted and unwanted radio
signals. So, filtering out the correct information is more
complicated then just looking for a peak in the electric field.

In a broad-band receiver the biggest contribution of the basic noise
level typically comes from narrow-band RFI transmitters. Hence, a
digital filter to cut these signals out -- which is currently only
employed in the post-processing, is a crucial step in the triggering
electronics. Moreover, the characteristics of the pulse itself need to
be considered as well. For a human eye it is quite simple to
distinguish a cosmic ray pulse from those generated by a passing 1970
Chevrolet. Hence pulse shape parameters need to be used in the
triggering. This will in any case require a bit more intelligence on
the trigger board, than with conventional experiments.  Such a
pulse-shape parameter search to implement self-triggering in hardware
is currently under way at LOPES$^{\rm STAR}$ showing some interesting
progress recently. These techniques will eventually be tested at Auger
and one can be hopeful that this last and crucial step will be
achieved in the not too distant future.

\section{Radio Pulses from the Moon} 
The main focus of this article was on the detection of radio emission
of air showers. However, we want to end with a few comments on the
prospects for radio emission from ultra-high energy cosmic rays with
upcoming radio telescopes, in particular with LOFAR. Here the
atmospheric detection will be implemented in the project --- with the
``transient buffer board'' being at the heart of this new
technique. It turns out that this buffer-board may also improve the
detection of cosmic rays hitting the moon.

\begin{figure*}
\begin{center}
\noindent
\includegraphics[width=0.6\textwidth]{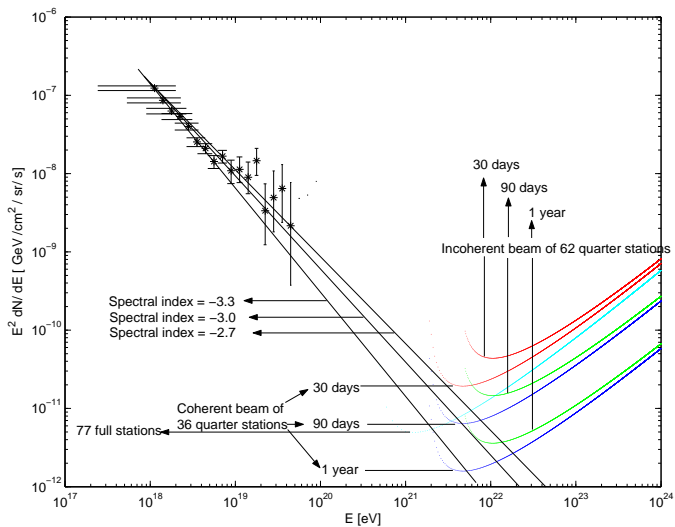}
\end{center}
\caption{Expected LOFAR sensitivities for detecting ultra-high energy
cosmic rays hitting the moon.``Full stations'' here relate to the
originally planned 96 tiles. 36 quarter stations are the currently
planned layout for the central LOFAR core. ``Coherent'' and ``incoherent''
relate to the two methods of combining the data from different stations. \label{figmoon}}
\end{figure*}
Scholten et al. \cite{ScholtenBacelarBraun2006} have shown that the
100-200 MHz range is ideal for detecting cosmic rays above $10^{20}$
eV hitting the lunar surface. In LOFAR any such event could be
detected in a beam formed towards the moon. This detection could be
used to trigger the buffer boards and download the raw data from all
LOFAR antennas. With the raw data of the individual antennas the exact
nature and origin of the pulse could be determined much more
precisely, if one uses some of the offline processing steps known from
the air shower detection.

The question is how sensitive is this technique? The originally
planned LOFAR would have been very favorable for this
\cite{ScholtenBacelarBraun2006}, however, due to the downsize of LOFAR the
sensitivity has decreased. K.~Singh has now recalculated the expected
sensitivity with the latest available station layout. The result is
shown in Fig.~\ref{figmoon}. With the lower sensitivity, the minimum
energy that can be detected has moved up above $10^{21}$ eV, which
will decrease the expected count rate.

For comparison we also show the latest Auger spectrum with an
extrapolation of the power law with three spectral indices through the
GZK cut-off.  For a powerlaw of $E^{-3}$ LOFAR could in principle
reach this extrapolation with a total observing time of 90 days, if a
large number of tied-array beams can be formed -- a mode that still
needs to be tested.

While strong evidence for a GZK cut-off above $10^{19.6}$ eV has been
seen by Auger, it cannot be excluded that there will be some recovery
of the spectrum from local sources or neutrinos. Hence, it is still
worth looking in this regime. Radio will probably be the only
technique that can deliver at least very meaningful upper limits in
this energy range.

Future expansions of LOFAR, such as the SKA, can improve these limits
even more, actually reaching down to the GZK cut-off with the clear
expectation of actual detections. This will then probably provide the
ultimate detection experiment for cosmic ray particles at the highest
energies we can ever measure.

\section{Conclusions}
The radio detection technique for cosmic particles has seen quite some
ups and downs in the last decades. Pronounced dead in the 1970's it has
now risen from the ashes. But is it here to stay? The chances are good
at least. 

First of all, we have made major progress in understanding the
emission mechanism. For radio in solid media (Cherenkov emission)
codes have been developed and accelerator experiments have been
performed, giving trust in the reality of the effect. For radio
emission from air showers (geosynchrotron) good Monte Carlo codes and
solid experimental verification are now available and more and more
details are being worked in.

The LOPES experiment, and in some areas also CODALEMA, has given us
already detailed information about the radio air shower properties and
performed a very useful cross-calibration between particle and radio
detectors.  Moreover, major experiments have embraced the
technique. The large LOFAR radio telescope has the cosmic ray
detection built in, the Auger collaboration is testing it, ANITA has
flown, and also the IceCube collaboration is seriously preparing radio
experiments.

A few issues still need to be solved: how does an optimal trigger system
look like for radio antennas? What is the optimal layout for a
large radio array? Nonetheless, there is a good chance that radio
detection will become common place over the next few years. 

It will be interesting to see in which direction this will develop.
The simulations for air showers indicate that the addition of radio may
increase energy resolution and directional accuracy. Also composition
information (through X$_{\rm max}$) seems to be encoded in the radio
signal. After the breakthrough of the hybrid technique with Auger,
maybe we will see ``tri-brid'' detectors in the future. More and
better data is almost always better in physics. If cosmic ray air
shower arrays are to continue in the next decades, they will likely
include radio antennas.  For the highest energy events, the new
generation of low-frequency radio telescopes provides hope that even
particles above $10^{21}$ eV could in principle be detected in the
future --- if they exist.


\end{document}